\documentclass[twoside,leqno,twocolumn]{article}
\usepackage[letterpaper]{geometry}
\usepackage{graphicx} 
\usepackage{amssymb}
\usepackage{mathtools} 
\usepackage{amsmath, amsthm, multirow, paralist}
\usepackage{ltexpprt}
\usepackage{multicol}
\usepackage{booktabs}
\usepackage{hyperref}
\usepackage{authblk}
\usepackage{hyperref}       
\usepackage{url}            
\usepackage{amsfonts}       
\usepackage{nicefrac}       
\usepackage{microtype}      
\usepackage{xcolor}         

\begin{document}
\title{\large DSAF: A Dual-Stage Adaptive Framework for Numerical Weather Prediction Downscaling}
\author{
    Pengwei Liu\textsuperscript{1}, 
    Wenwei Wang\textsuperscript{2}, 
    Bingqing Peng\textsuperscript{2}, 
    Binqing Wu\textsuperscript{1}, 
    Liang Sun\textsuperscript{2} \\
    \textsuperscript{1}Zhejiang University,  \textsuperscript{2}DAMO Academy, Alibaba Group\\
    \textsuperscript{1}\texttt{\{liupw,binqingwu\}@zju.edu.cn},
    \textsuperscript{2}\texttt{\{duoluo.www, pengbingqing.pbq, liang.sun\}@alibaba-inc.com}}

\date{}
\maketitle


\begin{abstract} \small\baselineskip=9pt 
While widely recognized as one of the most substantial weather forecasting methodologies, Numerical Weather Prediction (NWP) usually suffers from relatively coarse resolution and inevitable bias due to tempo-spatial discretization, physical parametrization process, and computation limitation. With the roaring growth of deep learning-based techniques, we propose the \textbf{D}ual-\textbf{S}tage \textbf{A}daptive \textbf{F}ramework (\textbf{DSAF}), a novel framework to address regional NWP downscaling and bias correction tasks. DSAF uniquely incorporates adaptive elements in its design to ensure a flexible response to evolving weather conditions. Specifically, NWP downscaling and correction are well-decoupled in the framework and can be applied independently, which strategically guides the optimization trajectory of the model. Utilizing a multi-task learning mechanism and an uncertainty-weighted loss function, DSAF facilitates balanced training across various weather factors. Additionally, our specifically designed attention-centric learnable module effectively integrates geographic information, proficiently managing complex interrelationships. Experimental validation on the ECMWF operational forecast (HRES) and reanalysis (ERA5) archive demonstrates DSAF's superior performance over existing state-of-the-art models and shows substantial improvements when existing models are augmented using our proposed modules. Code is publicly available at \href{https://github.com/pengwei07/DSAF}{https://github.com/pengwei07/DSAF}.
\end{abstract}

\section{Introduction}
With a century of evolution, Numerical Weather Prediction (NWP) has become one of the most substantial weather forecast methodologies, benefiting plentiful real-world applications such as climate research~\cite{hong2014dynamical}, disaster management~\cite{singh2022urban}, agricultural planning~\cite{fan2021medium}, and renewable energy forecasting~\cite{verzijlbergh2015improved}, etc. Basically, NWP is the numerical solution to massive physics-informed mathematical equations, which are initialized via data assimilation techniques and approximated by spatial and temporal discretization and physical parameterization processes. The roaring growth of computing power in the last few decades has contributed a lot to improving the accuracy and resolution of NWP. For example, the widely used HRES\footnote{https://confluence.ecmwf.int/display/FUG/HRES+-+High-Resolution+Forecast} is a single high-resolution forecast with horizontal resolution as 0.1$^{\circ}$ (9km) provided by European Centre for Medium-Range Weather Forecasts (ECMWF), and NCEP provides forecast with resolution as 0.25$^{\circ}$ (27km) from Global Forecast System (GFS) Model\footnote{https://www.ncei.noaa.gov/products/weather-climate-models/global-forecast}. However, in many real-world applications, such as wind power forecasting and event planning, people are more interested in the wind speed forecast at a wind farm or the precipitation forecast at a certain location. Thus, the coarse resolution forecasting provided by NWP is not satisfying.\par
The ultimate goal of NWP downscaling is to provide accurate weather forecasting with higher resolution. The first challenge of this task arises from the complicated spatial-temporal relationship among different weather factors (e.g., temperature, humidity, wind) in multiple locations~\cite{al2010review}. Secondly, the bias of current popular NWP is often observed due to the data quality and modeling process. Fig.~\ref{fig:DSAF} (a) and (b) show the comparison of correction, pure downscaling, and our approach. Thirdly, it is well observed that geographical information, such as proximity and topography, plays an important role in the interrelationship of climate data (as illustrated in Fig.~\ref{fig:DSAF} (d) and (e)). For example, adjacent locations with similar geographic characteristics share similar weather patterns, whereas disparate terrains can lead to significant meteorological variations~\cite{benavides2022review}. Therefore, it is crucial to incorporate the geographical data into NWP downscaling.\par  
The task of NWP downscaling for grid climate data bears a remarkable parallel to super-resolution (SR) tasks in the computer vision field, both aiming to establish a mapping from low-resolution inputs to high-resolution outputs. Deep Learning models, particularly Convolutional Neural Networks (CNNs), have demonstrated remarkable performance in image super-resolution~\cite{yang2019deep}. This shared objective has inspired a surge of deep learning-centric propositions for NWP downscaling~\cite{schultz2021can, yang2022statistical}. However, climate data differ from images in their spatial-temporal dependency and interaction among different weather factors. Thus, directly applying canonical SR models may lead to suboptimal results. Meanwhile, most existing NWP downscaling techniques neglect the coupled nature of correcting and downscaling, which frequently resort to end-to-end mapping~\cite{moiz2021evaluating, bochenek2022machine, harris2022generative}. Moreover, there is a conspicuous absence of systematic approaches to unravel the complex interrelationships among diverse weather factors~\cite{anupam2022nonlinear}, limiting models' representative capacities. The exploitation of geographic information would contribute to the downscaling process~\cite{marsh2023windmapper}, and how to better merge such information into the model is still under exploration.\par
To bridge this gap, we introduce \textbf{D}ual-\textbf{S}tage \textbf{A}daptive \textbf{F}ramework (\textbf{DSAF}), which innovatively disentangles the downscaling task into two distinct stages: correction and downscaling. Specifically, the first stage is a correction module taking the low-resolution input NWP data to calibrate the inherent bias. In the second stage, the corrected NWP data is then fed into a downscaling module to generate high-resolution NWP. Assisted by an attention-centric learnable module, DSAF incorporates geographic information, adeptly handling intricate relationships. To summarize, the main contributions of DSAF are:
\vspace{-1mm}
\begin{itemize}
\item A unique two-stage approach employing a channel-specific representation layer for correction and an innovative downscaling module based on a heterogeneous feature augmentation block, effectively enhancing detail and texture in high-resolution reconstructions.
\vspace{-1.5mm}
\item A holistic multi-task learning approach is employed to address the interplay of weather factors, which is achieved by integrating a weighted loss balancing method, a spatial-physical constraint, and a distinct channel separation strategy.
\vspace{-1.5mm}
\item Extensive experiments showcasing DSAF's superiority across weather datasets, particularly in wind speed, highlighting the effectiveness of our two-step approach and the importance of our correction-first, downscaling-next sequence.
\end{itemize}
\vspace{-2mm}
\section{Related Work}\label{RelatedWork}
\noindent\textbf{Super-Resolution Techniques in Climate Downscaling.} Due to the analogy between climate data and image data, the super-resolution techniques with deep learning, developed in computer vision, have been successfully applied for climate downscaling~\cite{yang2022statistical}. DeepSD~\cite{vandal2017deepsd} proposed a stacked SRCNN framework~\cite{dong2015image} for statistical downscaling of precipitation. FSRCNN~\cite{passarella2022reconstructing} lowers the computational cost by replacing a pre-processing step with a deconvolution layer at the end. Generative adversarial networks show promising performance in the downscaling task~\cite{stengel2020adversarial, izumi2022super, liu2022spatial}. Recently, topographical elevation information has been utilized to make super-resolution methods fitter in climate applications, and data augmentation modules are designed to enhance accuracy and robustness~\cite{park2022downscaling, rampal2022high}.\\
\noindent\textbf{Efficient Attention Mechanisms.} Climate data display strong spatial similarity, and the attention mechanism has been widely used in previous works. Multi-factor cross-attention module~\cite{jing2022attention}, terrain-guided attention network~\cite{liu2023statistical}, and residual dense channel attention block~\cite{xiang2022novel}, are proposed to capture global information and potential relationships between climate and geographic variables.\\
\noindent\textbf{Multi-task learning.} Muti-task learning has shown the ability to enhance model accuracy and efficiency~\cite{kendall2018multi, heydari2019softadapt}. Recently it has been adopted in climate fields, such as wind power prediction~\cite{fan2020m2gsnet}, lightning nowcasting~\cite{li2023convective}, and meteorological forecasting~\cite{ling2022multi}. 
\section{Methodology}\label{method}
\subsection{Problem Setting and Notations.} 
\begin{figure*}[t]
    \centering
    \vspace{-2mm}
    \includegraphics[width=\textwidth]{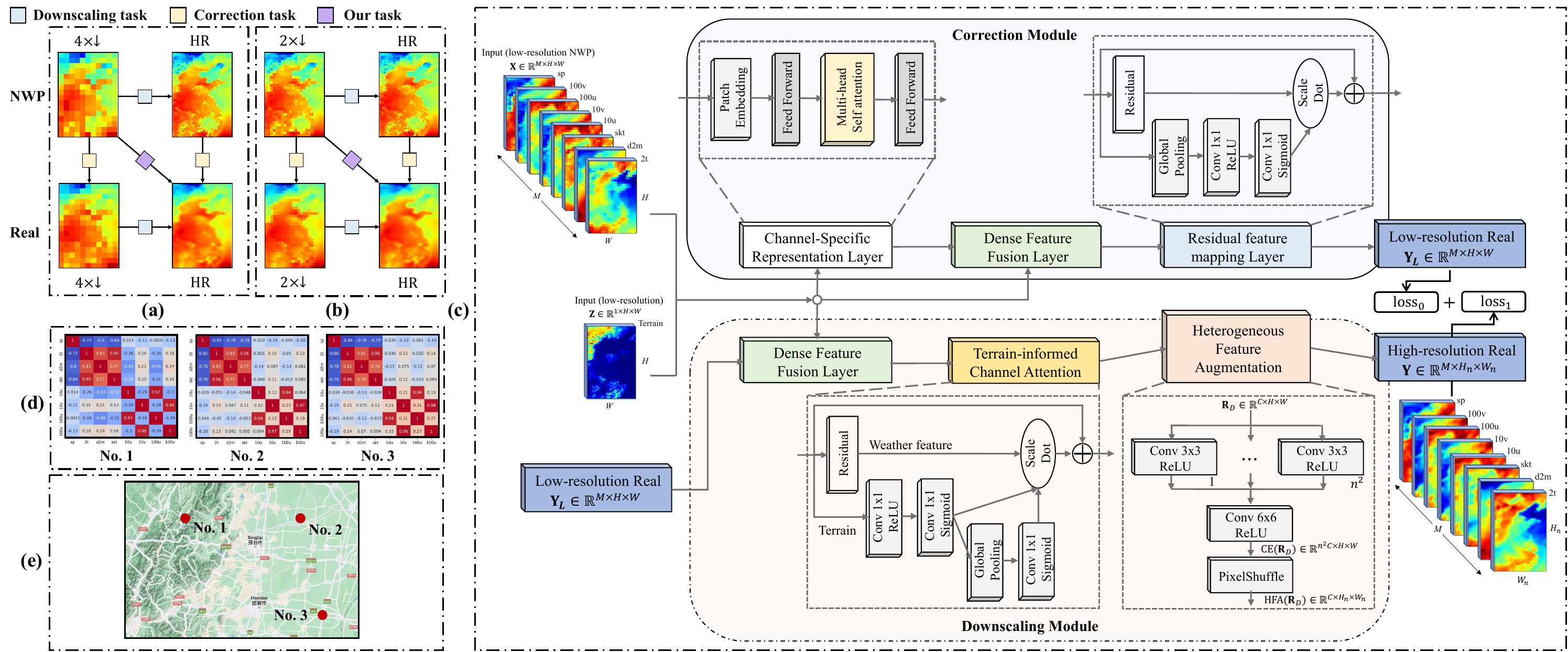}
    \vspace{-6mm}
    \caption{(a) and (b) Comparison of Correction, pure Downscaling and Our Tasks for 2m Temperature: Illustration of specific differences in $4\times$ and $2\times$ downscaling. NWP and Real denote HRES and ERA5 reanalysis data, respectively. HR signifies 0.25-degree resolution data, while $2\times\downarrow$ and $4\times\downarrow$ represent data at 0.5-degree and 1-degree resolutions, respectively. (c) Architectural overview of DSAF. (d) Heat maps of correlation matrices for diverse weather factors at Points No.1, No.2, and No.3. Warm colors indicate strong correlations, while cool colors represent weak ones. (e) Spatial and topographical characteristics of Points No.1, No.2, and No.3. Highlighting contrast between valley (Point No.1) and plain locations (Points No.2 and No.3).}
    \label{fig:DSAF}
    \vspace{-4mm}
\end{figure*}

As illustrated in Fig.~\ref{fig:DSAF} (a) and (b), DSAF is designed to perform NWP downscaling, which refines low-resolution forecasts into high-resolution outputs at any specified scale. Weather data inputs are represented as $\mathbf{X} \in \mathbb{R}^{M \times H \times W}$, where $M$ denotes the number of weather factors, $H$ and $W$ are the number of grids along longitudinal and latitudinal axes, respectively. Correspondingly, each grid is inherently linked with elevation data, denoted as $\mathbf{Z} \in \mathbb{R} ^ {1 \times H \times W}$.\par 
Formally, our task is to derive a function $f(\cdot)$, which takes low-resolution NWP data $\mathbf{X}$ and geographic information $\mathbf{Z}$ as input and generate high-resolution actual weather data $\mathbf{Y}$ at an arbitrary scale ($n$-scale). This process can be mathematically formulated as follows:
\begin{align}
\small{
f: \mathbb{R}^{M \times H \times W} \times \mathbb{R} ^ {1 \times H \times W} \rightarrow \mathbb{R}^{M \times H_n \times W_n},
\left[ \mathbf{X}; \mathbf{Z} \right] \xrightarrow{f(\cdot)} \mathbf{Y},
}
\end{align}
where $H_n$ and $W_n$ represent the longitudinal and latitudinal dimensions at the $n$-scale, $H_n = n \times H$ and $W_n = n \times W$.
\subsection{Model Overview.}
Our DSAF model is based on a dual-stage architecture incorporating a Correction Module and a Downscaling Module, as depicted in Fig. \ref{fig:DSAF} (c). Unlike conventional methodologies
~\cite{yu2021deep, passarella2022reconstructing, sharma2022resdeepd, jiang2022efficient} 
that undertake downscaling in a direct end-to-end approach, DSAF innovatively dissects the task into two distinct stages: correction and downscaling. This strategy is guided by the understanding that NWP models can be susceptible to inaccuracies due to inherent weather uncertainty and systematic bias in NWP products. To address these issues, DSAF introduces a correction stage prior to downscaling, which can effectively mitigate error propagation and enhance the overall precision of the downscaling procedure. 
The corrected data is then amalgamated with the original data via a residual structure, paving the way for the final downscaling stage executed through a dense block and feature fusion, ultimately yielding high-resolution data.
\subsection{Correction Module.}
Here we introduce key components in Correction Module: channel-specific representation layer (CRL), fusion layer, and mapping layer. \\
\noindent\textbf{Channel-Specific Representation Layer (CRL).} In light of the intricate interdependencies among different weather channels in NWP correction \cite{hu2021hybrid, laloyaux2022deep, ueyama2023radiative}, our approach employs a specialized representation layer to discern and leverage these relationships. For each weather channel $\mathbf{X}_i \in \mathbb{R}^{1 \times H \times W}, i={1,2,\cdots, M}$, we apply a patch embedding strategy to attenuate model complexity. In particular, the layer generates an embedding matrix $\mathbf{E}_{\text{all}}$ by $\mathrm{Concat}(\mathbf{E}_1,\cdots,\mathbf{E}_M)$, where $\mathbf{E}_i=\mathrm{MLP}(\mathrm{Flatten}\circ \mathrm{Patch}(\mathbf{X}_i)) \in \mathbb{R}^{F_{\text{p}} \times d}$, $\mathbf{E}_{\text{all}} \in \mathbb{R}^{F_P \times d}$, $F_P = M \times F_{p}$, and $d$ is the dimension of the embedding. Here $F_{p} = \frac{H}{p} \times \frac{W}{p}$, and $p$ is the dimension of the patch. Utilizing this channel-specific embedding, we extract information selectively from different representation subspaces using a Multi-Head Attention (MHA) block followed by a point-wise feed-forward network (FFN) as:
\begin{equation}
\small{
\left\{
\begin{aligned}
&\mathbf{A} = \mathrm{FFN}(\mathrm{MHA}(\mathbf{X})),\\
&\mathrm{MHA}(\mathbf{X}) = \left[\mathrm{head}_1, \mathrm{head}_2, \ldots, \mathrm{head}_h\right] \mathbf{W}^O, \\
&\mathrm{head}_i = \mathrm{Attention}\left( \mathbf{E}_{\text{all}}\mathbf{W}_i^Q, \mathbf{E}_{\text{all}} \mathbf{W}_i^K, \mathbf{E}_{\text{all}} \mathbf{W}_i^V\right),
\end{aligned}
\right.}
\vspace{-2mm}
\end{equation}
where $\mathbf{X} \in \mathbb{R}^{M \times H \times W}$ denotes $M$ weather factors, and $\mathbf{A} \in \mathbb{R}^{F_P \times p^2}$ is the output of CRL, whose dimension can be reshaped as $(M,H,W)$. The projection matrices $\mathbf{W}_i^Q$, $\mathbf{W}_i^K$, $\mathbf{W}_i^V$, and $\mathbf{W}^O$ are the respective learnable parameters for each attention head. And $\mathbf{W}_1$, $\mathbf{b}_1$, $\mathbf{W}_2$, and $\mathbf{b}_2$ are trainable parameters for FFN.\\
\noindent\textbf{Fusion Layer and Mapping Layer.} A deep fusion of the extracted features from the CRL is achieved through a dense layer, facilitating comprehensive information integration. The residual layer (Res) with channel attention (CA) is then applied, with its specific structure resembling that of the original RCAN~\cite{zhang2018image} while dropping the Batch Norm layer to boost high-frequency information flow. 
\subsection{Downscaling Module. }
Here we introduce key components in Downscaling Module, including terrain-informed channel attention (TCA) and heterogeneous feature augmentation. \\
\noindent\textbf{Terrain-informed Channel Attention.} Since NWP is highly related to terrain, which is static and in high resolution, we introduce the Terrain-informed Channel Attention (TCA) to integrate geographic information for NWP downscaling. Specifically, the TCA process is as follows:
\begin{equation}
\small{
\left\{
\begin{aligned}
& \mathbf{T}_{l} = \mathrm{TCA} (\mathbf{R}_{l-1},\mathbf{T}_{l-1}),\\
& \mathbf{R}_{l} = \mathrm{Res}\circ \mathrm{TCA} (\mathbf{R}_{l-1},\mathbf{T}_{l-1}) = \mathrm{Res}(\mathbf{R}_{l-1}) \odot \mathbf{T}_{l},\\
& \mathbf{T}_{0} = \mathbf{Z}, \mathbf{R}_{0} = \mathrm{Dense}(\mathrm{Concat}(\mathbf{X},\mathbf{Y}_L)), l = 1,\cdots,D,
\end{aligned}
\right.}
\end{equation}
where $\mathrm{TCA} (\mathbf{R}_{l-1},\mathbf{T}_{l-1}) = \mathbf{T}_{l-1} \odot \sigma(\mathrm{Conv}(\mathbf{T}_{l-1},\mathbf{R}_{l-1}))$, $\mathbf{R}_D$ is the output of the TCA process, $D$ denotes the number of layers of the residual block, $\mathbf{Z} \in \mathbb{R}^{1 \times H \times W}$ denotes terrain data and $\sigma$ is the sigmoid activation function. Through this nested structure, the geographic features can be deeply fused with the original data features to boost the model's performance further. \\
\noindent\textbf{Heterogeneous Feature Augmentation.} 
Upon completion of the TCA process, a diverse amalgamation of source features is achieved. Yet, our ultimate goal is to reconstruct high-resolution authentic data, $\mathbf{Y} \in \mathbb{R}^{M \times H_n \times W_n}$. Toward this aim, we employ a Heterogeneous Feature Augmentation (HFA) strategy. It has two key steps: channel expansion (CE) and pixel shuffle (PS). The CE step leverages the multifaceted nature of image features, wherein features corresponding to different levels are generated via $n\times n$ branches, with each branch learning relevant features. To enhance the smoothness between features, large convolution kernels are utilized for feature fusion. In the PS phase, which is an adaptation of the methodology proposed by Shi et al.~\cite{shi2016real}, we discern the intricate interconnections among features and generate high-resolution outputs. The pixel shuffle process expertly preserves minute details and textures, thereby resulting in images with superior clarity and realism. Specifically,
\begin{equation}
\small{
\left\{
\begin{aligned}
& \mathbf{Y} = \mathrm{Conv}(\mathrm{HFA}(\mathbf{R}_D)), \\
& \mathrm{HFA}(\mathbf{R}_D) = \mathrm{PS}\circ \mathrm{CE}(\mathbf{R}_D) = \mathrm{PS}(\mathrm{CE}(\mathbf{R}_D)),\\
& \mathrm{CE}(\mathbf{R}_D) 
= \mathrm{Conv}(\mathrm{Concat}(\overbrace{\mathrm{Conv}(\mathbf{R}_D), \cdots, \mathrm{Conv}(\mathbf{R}_D)}^{n^2}),\\
& \mathrm{PS}(\mathrm{CE}(\mathbf{R}_D)) = \{ Y_{c, i \times n + a, j\times n + b} \},\\
\end{aligned}
\right.
}
\end{equation}
where $\mathrm{CE}(\mathbf{R}_D) \in \mathbb{R}^{n^2\cdot C \times H \times W}$, $\mathrm{PS}(\mathrm{CE}(\mathbf{R}_D)) \in \mathbb{R}^{C \times H_n \times W_n}$ and $C$ is the channel number of $\mathbf{R}_D$. The term $Y_{c, i\times n + a, j\times n + b}$ denotes the pixel value at the index $(c, i\times n + a, j\times n + b)$ in $\mathrm{HFA}(\mathbf{R}_D)$, where $0 \leq a, b < n$, $0 \leq i < H, 0 \leq j < W$, and $0 \leq c < C$. This pixel value is mapped from the location value $(n^2\cdot c+n\cdot b+a, i, j)$ in $\mathrm{CE}(\mathbf{R}_D)$. Achieved by channel expansion, partitioning, and shuffling,  HFA effectively magnifies and reorganizes low-resolution data to form high-resolution outputs. The essence of HFA is reshaping a tensor to gain intricate patterns, providing the model with a richer understanding of multiscale data. As a result, it enables flexible downscaling of input data to any desired scale ($n$-scale).
\subsection{Loss Function. }
In our study, we tackle the complexities of diverse weather factors through a two-stage approach: correction and downscaling, similar to a multi-task learning setup. Balancing the loss terms of each weather factor is crucial, given their distinct characteristics within the data. With the DSAF, we merge the losses from both stages, resulting in a dual-task loss function. We also incorporate physical constraints linking wind speed and pressure weather conditions. The total loss function, $\mathrm{loss}_{\mathrm{DSAF}}$, is optimized to determine the best parameters $\theta$:
\vspace{-1mm}
\begin{equation}
\left\{
\begin{aligned}
& \mathrm{loss}_{\mathrm{DSAF}} = \mathrm{loss}_0 + \mathrm{loss}_1 + \lambda_{\text{reg}}\mathrm{loss}_{\text{reg}},\\
&\mathrm{loss}_{0} = \mathrm{MSE}(\mathbf{Y}_L, \mathbf{Y}_L^{*}),\\
&\mathrm{loss}_{1} = \mathrm{MSE}(\mathbf{Y}, \mathbf{Y}^{*}),
\end{aligned}
\right.
\end{equation}
where $\mathbf{Y}_L^{*}$ is the low-resolution observed data derived from real high-resolution data, $\mathbf{Y}^{*}$. The terms $\mathrm{loss}_{0}$ and $\mathrm{loss}_{1}$ represent correction and downscaling losses. Moreover, $\mathrm{loss}_{\text{reg}}$ is the spatial-physical constraint between wind speed ($\mathbf{u}$) and surface pressure ($\mathbf{p}$), following the general Poisson equation. The coefficient $\lambda_{\text{reg}}$ adjusts this loss constraint's weight.\\
\noindent\textbf{Integrating Uncertainty.} Adapting from \cite{kendall2018multi}, we employ the uncertainty weighing approach given the dual nature of weather factor correction and downscaling. For multiple distinct weather factors, the loss function is:
\begin{equation}
\small{
\vspace{-1mm}
\begin{aligned}
& \mathrm{loss}_i(\theta)= \sum_{j=1}^{M} \lambda_j^i L_j^i(\theta) = \sum_{j=1}^{M} \lambda_j^i \mathrm{MSE}(\mathbf{Y}_j^i, \hat{\mathbf{Y}}_j^i), i=0,1,
\end{aligned}
\vspace{-1mm}
}
\end{equation}
where $\lambda_j^i$ is the optimized loss hyperparameter. The terms represent channels of low-resolution input and its ground truth and high-resolution output with its corresponding truth, shedding light on individual weather factor behavior across resolutions.\\
\noindent\textbf{Spatial-physical Constraint.} The dynamic relationship between wind speed and pressure is an important meteorological research topic. Basically, as the pressure gradient intensifies, it often precipitates a surge in wind speed \cite{walker1990effects, kwak2007analysis}. This foundational relation can be represented by the general Poisson equation:
\begin{equation}\label{loss_reg}
\small{
Loss_{reg} = \nabla^2 p + \nabla \cdot (\mathbf{u} \cdot \nabla \mathbf{u}). }
\end{equation}
To further emphasize, in our practical observations, pressure's downscaling emerges as more intuitive compared to other meteorological variables. Such insights prompted us to harness pressure for refined wind speed correction. Driving the $Loss_{reg}$ towards zero accentuates the modeling of this relationship. 

Similar to physics encoding in ~\cite{rao2023encoding}, the core here is to introduce physical loss and fix the parameters of the convolution kernel with the calculation matrix, which corresponds to different order difference schemes. For example, we introduce a second-order central difference operator to approximate the second-order differential operator $\nabla^2$, corresponding to the $3\times3$ convolution kernel $K$, whose specific representation process is shown in Eq~\ref{fixed_conv}, where $h$ represents the step size, and we normalize it to $1$. In this work, we approximate the $\nabla$ operator of different orders in Eq~\ref{loss_reg} using the fourth-order different difference operator, corresponding to the $5\times5$ convolution kernel, thus represents $Loss_{reg}$.
\begin{equation}\label{fixed_conv}
\small{
\begin{aligned}
&\nabla^2 u(x, y)=\frac{\partial^2 u(x, y)}{\partial x^2}+\frac{\partial^2 u(x, y)}{\partial y^2} \\
& \approx \frac{u(x+h, y)-2 u(x, y)+u(x-h, y)}{h^2}\\
&+\frac{u(x, y+h)-2 u(x, y)+u(x, y-h)}{h^2} \\
& \approx u(x, y) \circledast \frac{1}{h^2}\left[\begin{array}{ccc}
0 & 1 & 0 \\
1&-4&1 \\
0 & 1 & 0
\end{array}\right] \coloneqq u(x, y) \circledast K .
\end{aligned}    
}
\end{equation}
\vspace{-6mm}
\section{Experiments}\label{experiments}
\subsection{Datasets.}
The datasets utilized in this study are derived from the European Centre for Medium-Range Weather Forecasts (ECMWF) operational forecast (HRES) and reanalysis (ERA5) archive. HRES represents a 10-day atmospheric model forecast, while ERA5 serves as a fifth-generation global atmospheric reanalysis, incorporating climate and weather observations. For regional NWP downscaling, we construct a real-world dataset called ``Huadong", covering the East China land and sea areas. In this dataset, HRES data is employed as the predictive data, while ERA5 reanalysis data serves as the ground truth.

\noindent\textbf{Dataset Details.} The Huadong dataset encompasses a latitude range from $26.8^\circ$N to $42.9^\circ$N and a longitude range from $112.6^\circ$E to $123.7^\circ$E. It comprises a grid of $64 \times 44$ cells, with each cell having a grid size of 0.25 degrees in both latitude and longitude. Notably, the Huadong dataset incorporates Digital Elevation Model (DEM) data to represent terrain information. The HRES and ERA5 data cover the period from January 3, 2020, to April 1, 2022, and include eight weather factors: surface pressure (`sp'), 2m temperature (`2t'), 2m dewpoint temperature (`d2m'), skin temperature (`skt'), 10m u component of wind (`10u'), 10m v component of wind (`10v'), 100m u component of wind (`100u'), and 100m v component of wind (`100v').

\noindent\textbf{Data Preprocessing.} We apply linear interpolation on the original HRES data, which has a grid size of 0.1 degrees, to harmonize it with the ERA5 data on a unified 0.25-degree resolution grid. Our experiments focus on $2\times$ and $4\times$ downscaling tasks, corresponding to resolutions of 0.5 degrees and 1 degree. To facilitate this, the 0.25-degree HRES data undergoes linear interpolation to generate the requisite 0.5-degree and 1-degree input data. Channel-wise normalization is performed for consistency and training efficiency, while the weather factor data for each channel is restored to its original spatial scale during loss function computation.
\subsection{Baselines.}
We benchmark our proposed DSAF model against the conventional Bicubic interpolation~\cite{bicubic1981} and five prominent models in the downscaling domain: FSRCNN~\cite{passarella2022reconstructing}, ResDeepD~\cite{sharma2022resdeepd}, EDSR~\cite{jiang2022efficient}, RCAN~\cite{yu2021deep}, and GINE~\cite{park2022downscaling}. 
FSRCNN is a widely recognized method in computer vision, leveraged for both downscaling and single-image super-resolution, which conducts feature mapping using multi-layer CNNs and executes upsampling via deconvolution layers.
ResDeepD and EDSR are built upon the ResNet architecture with different implementations: ResDeepD begins with an upsampling of the input to increase dimensions before proceeding to feature mapping via ResNet, while EDSR first conducts feature mapping using ResNet followed by upsampling. 
RCAN, also built on the ResNet architecture, differentiates itself by incorporating a global pooling layer for channel attention. 
GINE, an extension of RCAN with convolutional attention, integrates topographical elevation information into the downscaling process.
Note that these methods all perform downscaling in an end-to-end manner. 
To maintain the fairness of the experiment, We create optimized versions called C\_EDSR and C\_RCAN by adding a correction block, respectively. We also test a variant of our DSAF model, named D\_C\_DSAF, where downscaling is implemented preceding the correction process.
\begin{table*}[h]
\centering
\small
\setlength{\tabcolsep}{1.5pt}
\renewcommand{\arraystretch}{1}
\caption{Downscaling performance comparison of each weather factor on the ECMWF operational forecast (HRES) and reanalysis (ERA5) with RMSE. \textbf{Bold} and \underline{Underline} indicate the best and second best performance, respectively. $\Delta$ denotes the relative improvement between DSAF and baselines/Bicubic.}
\label{tab1}
\begin{tabular}{l|cc|cc|cc|cc|cc|cc|cc|cc}
\hline
Channel & \multicolumn{2}{c|}{sp} & \multicolumn{2}{c|}{2t} & \multicolumn{2}{c|}{d2m} & \multicolumn{2}{c|}{skt} & \multicolumn{2}{c|}{10u} & \multicolumn{2}{c|}{10v} & \multicolumn{2}{c|}{100u} & \multicolumn{2}{c}{100v} \\
\hline
Scale      & 2x & 4x & 2x & 4x & 2x & 4x & 2x & 4x & 2x & 4x & 2x & 4x & 2x & 4x & 2x & 4x\\
\hline
Bicubic    
&0.82&1.22&1.71&1.84&1.99&2.06&1.61&1.85&1.21&1.27&1.28&1.36&1.68&1.76&1.78&1.87 \\
FSRCNN &0.43 &0.56 &1.20 &1.23 &1.24 &1.39 &1.29 &1.33 &1.09 &1.12 &1.07 &1.09 &1.19 &1.20 &1.23 &1.27\\
ResDeepD &0.30 &0.32 &1.18 &1.20 &1.26 &1.29 &1.20 &1.23 &1.00 &1.07 &0.97 &1.01 &1.09 & 1.11 &1.08 &1.11\\
EDSR     
& 0.46 & 0.55 & 1.21 & 1.27 & 1.30  & 1.38 & 1.24 & 1.39 & 1.02 & 1.11 & 1.00 & 1.11 & 1.14 & 1.21 & 1.17 & 1.22 \\
C\_EDSR     
& 0.22 & 0.18 & 1.19 & 1.11 & 1.26 & 1.20  & 1.12 & 1.10  & 0.75 & 0.74 & 0.78 & \underline{0.73} & 1.02 & 1.04 & 1.07 & 0.98 \\
RCAN       
& 0.57 & 0.61 & 1.22 & 1.26 & 1.29 & 1.38 & 1.28 & 1.31 & 1.03 & 1.10  & 1.00 & 1.10  & 1.16 & 1.19 & 1.19 & 1.20  \\
C\_RCAN     
& \underline{0.20}  & \underline{0.190} & \underline{1.00} & \underline{0.99} & 1.13 & 1.09 & \underline{0.99} & \underline{0.98} & 0.81 & \underline{0.72} & \underline{0.76} & 0.74 & \underline{1.00} & \underline{1.00} & 0.98 & \underline{0.95} \\ 
GINE&0.38&0.40&1.02&1.09&\underline{1.03}&\underline{1.07}&1.00&1.10&\underline{0.73}&0.81&0.79&0.87&1.03&1.05&\underline{0.97}&1.13\\
D\_C\_DSAF &0.21 &0.191 &1.01 &1.00 &1.15 &1.10 &1.02 &1.01 &0.78 &0.73 &0.77 &0.76 &1.01 &1.02 &1.03 &1.01\\
DSAF (ours)   
&\textbf{0.17}&\textbf{0.15}
&\textbf{0.96}&\textbf{0.95}
&\textbf{1.02}&\textbf{1.01}
&\textbf{0.89}&\textbf{0.88}
&\textbf{0.66}&\textbf{0.65}
&\textbf{0.68}&\textbf{0.65}
&\textbf{0.92}&\textbf{0.89}
&\textbf{0.86}&\textbf{0.85}\\
\hline
$\Delta$ Best  
& 15.0\% & 21.1\% 
& 4.0\% & 4.0\% 
& 1.0\% & 5.6\% 
& 10.1\% & 10.2\% 
& 9.6\% & 9.7\% 
& 10.5\% & 11.0\% 
& 8.0\% & 11.0\% 
& 11.3\% & 10.5\% \\
$\Delta$ Bicubic 
& 79.3\% & 87.7\% 
& 43.9\% & 48.4\% 
& 48.7\% & 51.0\% 
& 44.7\% & 52.4\% 
& 45.5\% & 48.8\% 
& 46.9\% & 52.2\% 
& 45.2\% & 49.4\% 
& 51.7\% & 54.6\% \\
\hline
\end{tabular}
\vspace{-4mm}
\end{table*}
\subsection{Experimental Settings}
\textbf{Metrics.} The Root Mean Squared Error (RMSE) is adopted to measure the difference between model prediction and the actual values for model evaluation. Formally, we evaluate the model's performance separately for each weather factor by $\mathrm{L}_{j} = \frac{1}{N}\sum_i^{N}\mathrm{RMSE}(\mathbf{Y}_j^i,\hat{\mathbf{Y}}_{j}^i),$
where $\mathrm{L}_{j}$ denotes the performance of the model for the $j^{th}$ weather factor, $N$ represents the number of samples in test dataset for the $j^{th}$ weather factor, and $\mathbf{Y}_j^i$ and $\hat{\mathbf{Y}}_{j}^i$ refer to the $j^{th}$ channel of $\mathbf{Y}$ and $\mathbf{Y}^{*}$, respectively.

\noindent\textbf{Implementation Details.} 
Our models are implemented on a single 32GB Tesla V100 GPU using the PyTorch framework. We've opted for the Adam optimizer \cite{kingma2014adam}, with a starting learning rate of $0.0005$. Throughout our training, which spans up to 200 epochs, we're watchful of the validation loss and stop the process if it remains stagnant over 15 epochs. Key hyperparameters include $\lambda_{\text{reg}}$ for the loss term, $L$ for the number of residual layers, $h$ for the heads in MHA, $p$ as the patch dimension, $num\_d$ for the layers in the dense block, and $g$ to monitor the model growth rate. Downscaling segment parameters are $D$, $num\_d1$, $g1$, and $scale$ for the number of residual layers in TCA, dense layer number and growth rate, and downscaling scale. For our configuration, parameters are set as $\lambda_{\text{reg}}=0.1$, $L=D=3$, $num\_d = num\_d1 = 3$, $g = g1 = 16$. Task-wise, the $2\times$ setting adopts $h=4$, $p=2$, and $scale=2$, whereas the $4\times$ task uses $h=1$, $p=1$, and $scale=4$. It's also worth noting that the channel count stands at $128$ for the residual blocks in both modules, and our model benefits from an adaptive average pooling layer that facilitates global pooling.

\noindent\textbf{Channel Separation.} To enhance our model's performance and efficiency, we adopt a channel separation strategy based on unique weather factor correlations. As shown in Fig.~\ref{fig:DSAF} (d), factors with significant correlations, such as temperature and wind, are clustered as unified entities during training, and a channel separation procedure is applied thereafter.
\subsection{Main Results.}
The comparison between DSAF and baseline models is summarized in Table \ref{tab1}, and we highlight the corresponding observations as follows. 

\noindent\textbf{DSAF's Performance Superiority.} (1) Our proposed DSAF model exhibits significant proficiency in NWP downscaling, boasting an average performance enhancement surpassing 50\% when compared with the Bicubic model. (2) DSAF consistently outperforms all other models across all weather factor datasets, with a particular superiority in wind speed. This can be attributed to DSAF's effective capturing and utilization of the spatial correlations among different weather channels, as well as its application of HFA to bolster the detailed texture in super-resolution reconstruction. Thus, DSAF demonstrates commendable performance in reconstructing wind speed weather factors due to their strong continuity.

\noindent\textbf{Effectiveness of the Two-Stage Method.} The proposed two-stage method’s effectiveness is confirmed through the notable performance improvement of C\_EDSR and C\_RCAN compared to their original counterparts. This improvement also justifies our approach: prioritizing NWP data correction before downscaling within real observations at a lower resolution aids in mitigating model error accumulation, subsequently enhancing the overall accuracy of the downscaling process.

\noindent\textbf{Impact of Task Sequencing and Two-Stage Design.} (1) In the realm of end-to-end models, the $2\times$ task consistently outperforms the $4\times$ task. Conversely, in the context of the two-step models, the $4\times$ task showcases superior performance, affirming the significance of our two-stage design for multi-resolution downscaling. (2) Furthermore, a comparison between the DSAF model and the D\_C\_DSAF method underscores the substantial influence of task sequencing on model performance, reiterating the efficacy of our advocated correction-first, downscaling-next approach.
\begin{table*}[t]
\setlength{\tabcolsep}{3.7pt}
\renewcommand{\arraystretch}{1}
\centering
\caption{Comparison of different ways of Multi-task learning and Ablation study.}
\label{tab2}
\small
\begin{tabular}{l|cc|cc|cc|cc|cc|cc|cc|cc}
\hline
Channel & \multicolumn{2}{c|}{sp} & \multicolumn{2}{c|}{2t} & \multicolumn{2}{c|}{dew} & \multicolumn{2}{c|}{skin} & \multicolumn{2}{c|}{10u} & \multicolumn{2}{c|}{10v} & \multicolumn{2}{c|}{100u} & \multicolumn{2}{c}{100v} \\
\hline
Scale      & 2x & 4x & 2x & 4x & 2x & 4x & 2x & 4x & 2x & 4x & 2x & 4x & 2x & 4x & 2x & 4x\\
\hline
Base\_DSAF     
&0.82&1.22&1.71&1.84&1.99&2.06&1.61&1.85&1.21&1.27&1.28&1.36&1.67&1.76&1.78&1.87 \\
DeConv\_DSAF 
&0.41&0.40&1.29&1.09&1.30&1.16&1.11&1.00&0.82&0.84&0.99&0.83&1.07&1.14&1.14&1.13 \\
no\_TCA         
&0.41&0.39&1.23&1.08&1.26&1.16&1.10&1.01&0.79&0.76&0.89&0.84&1.02&1.06&1.00&1.00 \\\hline
single       
&0.37&0.33&1.18&1.08&1.24&1.15&1.09&1.00&0.79&0.76&0.91&0.79&1.05&1.10&1.09&1.05 \\\hline
Average      
&0.41&0.36&1.17&1.12&1.22&1.18&1.15&1.08&0.87&0.83&0.89&0.85&1.11&1.10&1.13&1.17 \\
Adaptive     
&0.38&0.36&1.14&1.10&1.20&1.17&1.09&1.01&0.76&0.75&0.78&0.83&1.05&1.06&1.09&1.22 \\\hline
no-Phy
&0.18&0.16&0.96&0.96&1.03&1.01&0.90&0.88&0.73&0.72&0.75&0.72&1.00&0.99&0.94&0.92 \\\hline
DSAF (ours)   
&\textbf{0.17}&\textbf{0.15}
&\textbf{0.96}&\textbf{0.95}
&\textbf{1.02}&\textbf{1.01}
&\textbf{0.89}&\textbf{0.88}
&\textbf{0.66}&\textbf{0.65}
&\textbf{0.68}&\textbf{0.65}
&\textbf{0.92}&\textbf{0.89}
&\textbf{0.86}&\textbf{0.85}\\
\hline
\end{tabular}
\end{table*}

\begin{figure*}[h]
\centering
\vspace{-2mm}
   \begin{minipage}{0.476\textwidth}
        \centering
        \includegraphics[width=\textwidth, trim=3.5cm 3.2cm 0.4cm 3.1cm, clip]{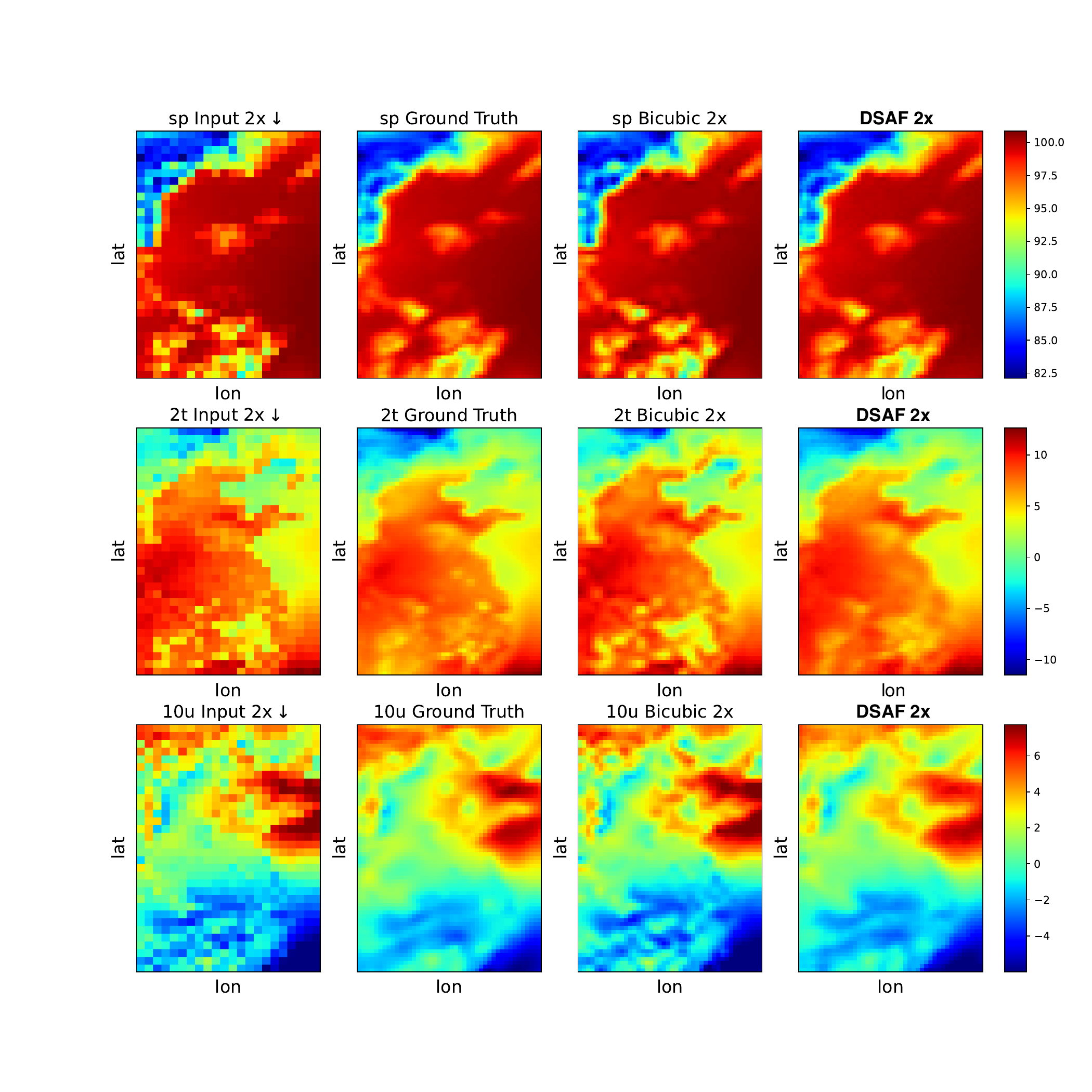}
        \vspace{-6mm}
        \caption{Visualization of $2\times$ Downscaling.}
        \label{fig:2xresults}
        \vspace{-2mm}
    \end{minipage}
        \qquad
    \begin{minipage}{0.476\textwidth}
        \centering
        \includegraphics[width=\textwidth, trim=3.5cm 3.2cm 0.4cm 3.1cm, clip]{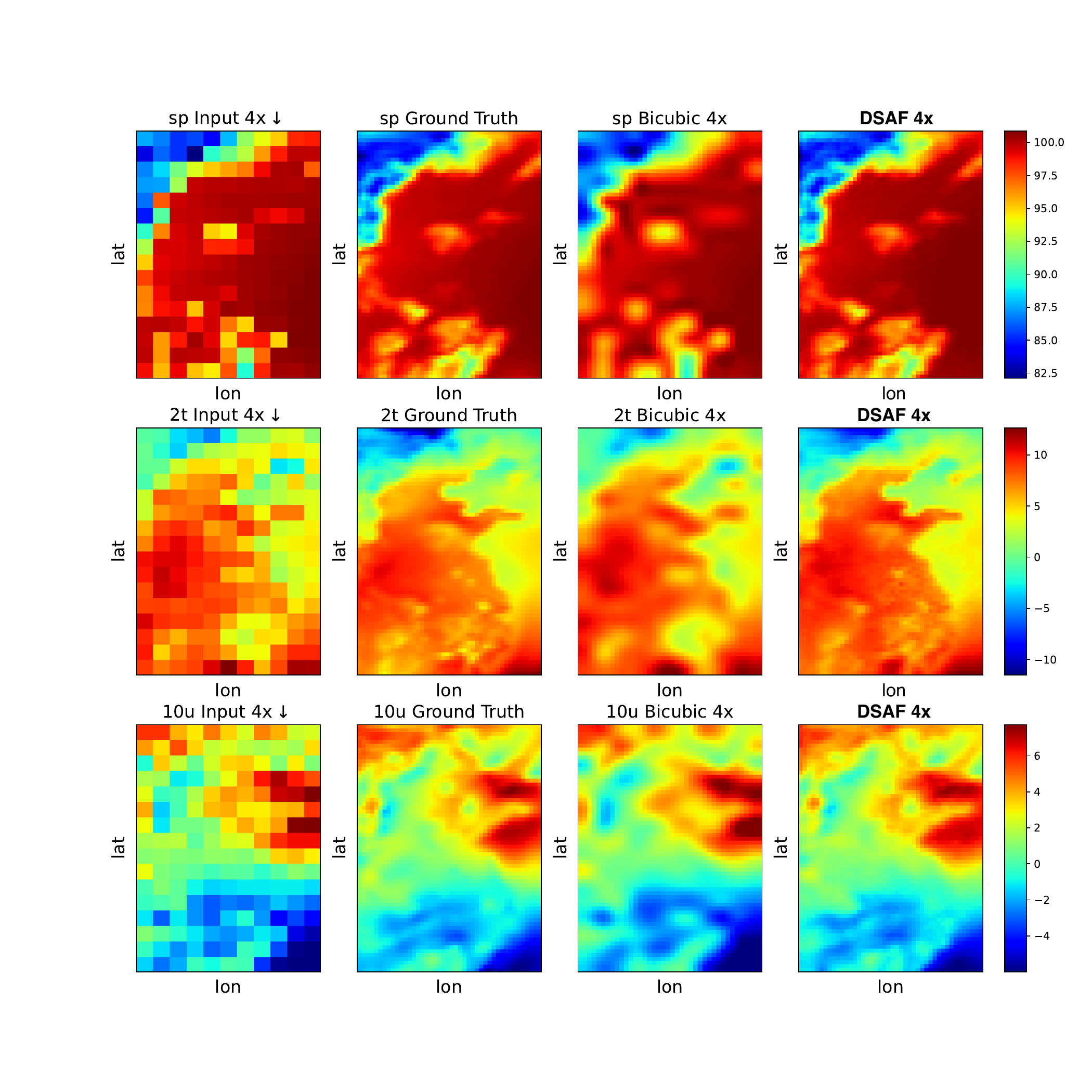}
        \vspace{-6mm}
        \caption{Visualization of $4\times$ Downscaling.}
        \label{fig:4xresults}
        \vspace{-2mm}
 \end{minipage}
\vspace{-2mm}
\end{figure*}
\subsection{Model Evaluation.}
Here we study the multi-task learning approaches, ablation studies, and visualization. 

\noindent\textbf{Multi-task Learning Approaches.}
As shown in Table \ref{tab2}, we scrutinize three multi-task learning strategies. `Average' calculates the mean loss per weather factor, while `Adaptive' integrates all eight weather factors into the model, balancing their loss via an uncertainty weighing method. Our chosen strategy, `Channel separation+Adaptive', segregates weather factors into two categories, balancing the loss within the same class using uncertainty weighing. Rigorous testing on $2\times$ and $4\times$ tasks across eight weather factors confirms the superior performance of our `Channel separation+Adaptive' approach.

\noindent\textbf{Ablation Studies.} Table \ref{tab2} presents our ablation studies conducted on various weather factors using four distinct model variants: Base\_DSAF, DeConv\_model, no\_TCA, and `single'. Specifically, Base\_DSAF incorporates only the dense and residual layers within its correction and downscaling blocks. In contrast, DeConv\_model replaces the HFA block with a deconvolution architecture to generate high-resolution images. The no\_TCA variant omits terrain data by removing the terrain branch from DSAF's downscaling block's residual layer. Meanwhile, the 'single' configuration sidesteps the multi-task strategy, employing a dedicated correction and downscaling mechanism for each weather factor with an isolated input channel. In addition to these, we also executed a 'no-phy' test where the Spatial-physical Constraint isn't factored into the loss function, retaining only loss\_1 and loss\_2. Our experiments confirm the effectiveness of our architecture, especially the CRL and HFA block, enabling effective capture and utilization of spatial correlation among weather channels for detailed super-resolution reconstruction. The multi-task learning strategy further enhances our model's representational capacity, enabling the handling of multiple heterogeneous weather factors and balancing loss effectively.\\
\noindent\textbf{Visualization.} Fig. \ref{fig:2xresults} and Fig. \ref{fig:4xresults} plot the downscaling outcomes at $2\times$ and $4\times$ scales, respectively, using three representative weather factors `sp', `2t', and `10u' as examples. Each figure features four columns: the first showcases the low-resolution model input; the second displays the high-resolution ground truth; the third presents high-resolution data obtained via the Bicubic method applied to the low-resolution input; and the fourth reveals the results from our model's $2\times$ and $4\times$ downscaling operations. Significantly, within an identical model structure, the $4\times$ downscaling task surpasses the $2\times$ task. This validates our proposition that lower-resolution data correction facilitates effective high-resolution weather data reconstruction. The reconstruction quality notably enhances as deviation diminishes, especially when the input data closely matches actual data. These insights underscore the critical role of the correction branch in our model, reinforcing our design strategy.
\section{Conclusion}
In this paper, we propose DSAF, a novel framework for NWP that addresses the coupled task of correction and downscaling, heterogeneity among different weather factors, and spatial correlations within the weather data. The model's dual-stage structure, multi-task learning strategy, and the incorporation of spatial similarity have shown their effectiveness in extensive experiments. Despite these advancements, several interesting directions remain for future work. 
More sophisticated techniques such as Fourier Neural Operators \cite{li2020fourier} could be incorporated to better model the interdependencies among different weather factors and their correlations with the terrain. On the other hand, we plan to apply the downscaled NWP to downstream applications such as wind power forecasting.  
\bibliographystyle{siam}
\bibliography{mybib}
\end{document}